\documentclass[prb,twocolumn,showpacs,superscriptaddress,floatfix]{revtex4}
\usepackage{graphicx}
\newcommand{\be}{\begin{equation}}
\newcommand{\ee}{\end{equation}}
\newcommand{\bea}{\begin{eqnarray}}
\newcommand{\eea}{\end{eqnarray}}

\newcommand{\f}{\frac}

\newcommand{\Gbar}{\bar{G}}
\newcommand{\bGm}{{\bf G_A}}
\newcommand{\bGbar}{{\bf {\bar{G}}}}

\newcommand{\bGi}{{\bf G^{i}}}
\newcommand{\bSigmainv}{{\bf \Sigma^{-1}}}

\begin{document}

\title{Using the Average Spectrum Method to extract Dynamics from Quantum Monte Carlo simulations} 

\author{Olav  F. Sylju{\aa}sen}

\affiliation{NORDITA, Blegdamsvej 17, DK-2100 Copenhagen {\O}, Denmark; \\ Department of Physics, University of Oslo, P.~O.~Box 1048 Blindern, N-0316 Oslo, Norway}

\date{\today}


\preprint{NORDITA-2007-18}

\begin{abstract}
We apply the Average Spectrum Method to the problem of getting the excitation spectrum from imaginary-time quantum Monte Carlo simulations. We show that with high quality QMC data this method reproduces the dominant spectral features very well. It is also capable of giving information on the spectrum in regions dominated by the many-particle continuum of excitations.  
\end{abstract}

\maketitle

\section{introduction \label{introduction}}
Quantum Monte Carlo simulations (QMC) has become the method of choice for studying large equilibrium quantum many-body systems without approximations. While it is possible to obtain thermodynamic and static properties to a high degree of accuracy with QMC, it is almost a paradox that estimates for the excitation spectrum and the equilibrium dynamics are typically obtained with much less accuracy. The technical reason for this is that QMC is invariably formulated in {\em imaginary} instead of real time. This is not just a matter of choice, in fact the imaginary time formulation is necessary to avoid crucial sign problems which would ruin the statistical accuracy of the method. The difficulty in obtaining the dynamics lies in transforming imaginary time correlation functions back to real time. This ``Wick rotation'' is easily carried out when an {\em analytic} expression of the imaginary time correlation function is known. However, when only numerical data and their associated error bars are available, as in QMC, it is well known that the direct transformation is ill-defined and very sensitive to the errors. 
 
The common way to deal with this problem is to treat the transformation to real frequencies as a problem in data analysis where the imaginary time QMC result plays the role of the data and the real frequency spectral function is the sought-after model underlying the data. The data analysis problem is approached using Bayesian statistics which aims at identifying probabilities for different spectral functions that can account for the observed imaginary time data. In finding the best spectral function it is important that the spectral function not only fits the data well, but also that it is consistent with prior knowledge about which types of spectral functions are permissible. The Bayesian statistical framework is well suited for this as both prior knowledge and data-fitting are taken into account. 

Although not often coined in the Bayesian language, the procedure of fitting certain specific functional forms to the imaginary time data, is an example of Bayesian analysis where the prior probability distribution assigns equal probabilities to spectral functions of the specific functional form and the fitting procedure selects the best functional parameters. However, fitting to a certain class of functions assumes a rather high degree of prior knowledge. While such knowledge should be used whenever available it is not so common that one actually knows the exact functional form of the spectral function a priori.   

It is more often the case that one does not know the actual shape of the spectral function, but only knows certain sum rules and physical requirements such as real-valuedness and positivity. One should then prefer a prior probability distribution that takes only into account the prior knowledge and do not make extra assumptions. Such a maximally non-committal prior probability distribution is gotten by maximizing the entropy of the distribution under constraints coming from the specific a priori knowledge\cite{Shannon,JaynesBook}. In carrying out such a maximization it is important to consider the correct space to perform it in. A probability distribution of spectral functions is clearly multidimensional. Yet it is customary to treat the spectral function itself as a one dimensional probability distribution and choose a prior probability distribution that gives a high probability to spectral functions having a large entropy\cite{MaxEntReview}. Thus instead of maximizing the entropy of the multidimensional {\em probability distribution} of spectral functions, the entropy of the spectral function itself is maximized. The latter is not the maximally non-committal probability distribution taking into account only positivity and sum-rules. In fact, to arrive at this socalled entropic prior involves additional assumptions\cite{GullSkilling}, which applicability to the problem at hand is questionable, and often one finds that methods using the entropic prior gives too broad spectral features. In this article we favor the use of another less constraining prior which reflects explicitly what a priori information is included. 

In this article we use the Average Spectrum Method (ASM), first proposed in Ref.~\onlinecite{White}, where the posterior probability distribution is composed of a likelihood function and a weakly constraining prior. In the ASM the final spectrum is obtained as the average spectrum over this a posteriori probability distribution, thus the name ASM. We show examples of its use in getting not only the dominant features of the excitation spectra of quantum many-body models but also to a certain extent subdominant features.

This article is structured as follows: In section \ref{Bayesian_method} the Bayesian method is reviewed and the prior probability distribution is presented. The ASM is explained in Section \ref{Sampling_the_posterior}, and the particular Monte Carlo implementation of it used in this article is described in Section \ref{Monte_Carlo_implementation}. In section \ref{Applications} the ASM is applied to several different quantum spin systems. The article ends with a summary.

\section{Bayesian method \label{Bayesian_method}} 

The equilibrium dynamics of a physical system is characterized by the spectral function $A(\omega)$ which is real and non-negative. However, in QMC what is typically obtained is an imaginary time correlation function $G(\tau)$ which is related to the spectral function as 
\be \label{fundamentalrelation}
       G(\tau) = \int d\omega K(\tau,\omega) A(\omega) 
\ee
where the kernel of the transform $K(\tau,\omega)$ takes on different forms depending on whether the operators in the measured correlation function are fermionic or bosonic. 
In order to make the discussion definite and practical we will model the spectral function as a collection of $N$ delta-functions on a frequency grid ${\omega_i}$
\be
       A(\omega) = \sum_{i=1}^{N} A_{\omega_i} \delta (\omega - \omega_i),
\ee
where all $A_{\omega_i}$ are positive or zero.
We will take a regularly spaced frequency grid such that $\omega_i$ is independent of $i$ up to a frequency cutoff $\omega_{\rm max}$ which is chosen to be several times the bandwidth of the system in question. 
This choice of frequency grid is not necessarily an optimal choice as it might be more effective to choose a finer grid where the spectral function is varying most. However, in the absence of such a priori information the choice of a uniform grid up to a large cutoff value is reasonable.

Furthermore we will assume that $G(\tau)$ is obtained in QMC simulations and recorded at discrete imaginary times $\tau$. With this Eq.~(\ref{fundamentalrelation}) takes the form
\be \label{discreteeq}
       G_\tau = \sum_i K_{\tau,\omega_i} A_{\omega_i}.
\ee
The goal is to invert this relation. This is an ill-posed problem because of the near-zero eigenvalues of the kernel and therefore very sensitive to statistical errors of $G_\tau$.

In the Bayesian approach one instead attempts to find the {\em probability} of a particular spectral function $A$ given the QMC imaginary data $G$ and prior knowledge. This, {\em posterior} probability $P(A|G)$, can be expressed using Bayes theorem as
\be \label{Bayes}
   P(A|G) \propto P(G|A) P(A)
\ee
where $P(G|A)$ is the likelihood that the QMC data turns out to be $G$ given a particular spectral function $A$, and $P(A)$ is the prior probability distribution of the spectral function. The prior probability distribution encodes the knowledge we have about the spectral function $A$ before any QMC data is obtained.
   
Eq.~\ref{Bayes} raises the question of how to concretely express the prior probability distribution $P(A)$. We will use the following expression
\be \label{generalprior}
   P(A) \propto \delta(\sum_i K_{0 \omega_i} A_{\omega_i} - G_0) \Pi_i \Theta(A_{\omega_i})
\ee
which assigns equal probabilities to all spectral functions that satisfy the non-negativity requirement ($A_{\omega_i} \ge 0$) and the zero moment sum rule $\sum_i K_{0 \omega_i} A_{\omega_i} = G_0$.
In Eq.~\ref{generalprior}  $\Theta(x)=1$ for $x \geq 0$ and zero otherwise. The product of $\Theta$-functions incorporates the knowledge that all spectral components must be non-negative, and the $\delta$-function constrains the spectra to obey the zero-moment sum rule. Higher order sum rules can be implemented by multiplying by more $\delta$-functions. This prior probability distribution is the probability distribution having the highest entropy consistent with the requirement of the non-negativity constraint and the zeroth moment sum rule. It is therefore not a very selective probability distribution as it gives the same probability to any spectral function that satisfy the sum rule and is non-negative.

\section{the Average Spectrum Method \label{Sampling_the_posterior}}
Given the weak discriminating nature of the prior, Eq.~(\ref{generalprior}), it is not a good idea to pick as the final answer the spectral function that maximizes the posterior probability distribution. It is rather obvious that the spectrum obtained in that way will over-fit the data in the sense that it also will fit the noise. Instead we will pick as the final answer the {\em average} spectral function, obtained by averaging over the posterior distribution\cite{White}. Thus we will compute
\be
     \bar{A} = \int dA A P(A|G)/\int dA P(A|G).
\ee
The averaging procedure itself will protect against over-fitting the data. The averaging procedure tends to smooth out the spectral function, and, in fact, it has been shown that when the average is carried out within the mean field approximation the result is identical to the classic MaxEnt result\cite{Beach}. However, in general the methods yield different results.

It is appropriate here to compare and contrast the ASM to the more commonly used MaxEnt methods\cite{MaxEntReview}. The methods differ in that in MaxEnt methods an entropic prior is assumed for the spectral function and not the prior specified in Eq.~\ref{generalprior}. In MaxEnt methods the entropic prior is multiplied by a factor $\alpha$ which determines how much influence it has compared to the likelihood-function. Different MaxEnt methods differ in how the final answer for the spectral function is arrived at. In the classic MaxEnt method the probability distribution for the parameter $\alpha$, $\pi(\alpha)$, is determined by Bayesian inference and the final answer is picked as the spectral function corresponding to the value of $\alpha$ that maximizes this probability distribution. Bryan's MaxEnt method\cite{Bryan}, on the other hand, is more similar to the ASM method as there the final spectrum is obtained by {\em averaging} the different spectral functions obtained at different values of $\alpha$ over $\pi(\alpha)$. This can either be done by computing $\pi(\alpha)$ directly for a range of $\alpha$'s and averaging their spectra, or by using a Monte Carlo procedure as shown in Ref.~\onlinecite{Boninsegni}.

Taking the average as the final answer is appropriate when the posterior probability has a single prominent peak. However, when there are more peaks the meaning of the average becomes more questionable. In order to detect such multiple peak situations one can focus on a few spectral features and make histograms of these according to the posterior probability distribution, and check for multiple peaks in these histograms. 

The averaging procedure can be efficiently carried out using Monte Carlo methods. In the context of getting dynamics from QMC this approach is known as the Average Spectrum Method\cite{White}, or Stochastic continuation\cite{Stocon}, but it is also used for data analysis in many other fields, see for instance Refs.~\onlinecite{Tarantola} and \onlinecite{Genetics}, where it is generally known as Markov Chain Monte Carlo methods.

To compute the posterior probability $P(A|G)$ we also need the likelihood function $P(G|A)$. Assuming that the imaginary time data is distributed as Gaussians with covariance matrix ${\bf \Sigma}$, the likelihood function $P(G|A)$ is
\be
   P(G | A) \sim e^{-\f{1}{2} Tr \sum_i \left( \bGi -\bGm \right)^T \bSigmainv \left( \bGi - \bGm \right) }
\ee
where we have denoted by ${\bf G^{i}}$ a vector of imaginary time values $G^i_\tau$ that is the average result of the i'th bin of QMC data containing $M$ measurements. 
The assumption of having Gaussian data should be good for large amount of data, however this assumption should always be checked for instance by monitoring skewness and kurtosis. 
Similarly we denote by ${\bf G_A}$ a vector with components
\be
    G_{A \tau} = \sum_{j} K_{\tau \omega_j} A_{\omega_j}
\ee
coming from a particular spectral function $A_\omega$.
In total there are $n$ bins of QMC data, and for large $n$, ${\bf \Sigma}$ can be approximated by the measured covariance matrix having components 
\be
   \Sigma_{kl} \approx \f{1}{n-1} \sum_i \left( G^{i}_{\tau_k} - \Gbar_{\tau_k} \right) 
                            \left( G^{i}_{\tau_l} - \Gbar_{\tau_l} \right) 
\ee
where we have denoted by an over-bar the total mean of the QMC data
\be
    \bGbar = \f{1}{n} \sum_i \bGi.
\ee
It is useful to express the posterior probability in terms of this total mean.
Using the cyclic property of the trace the exponent can be written as
\bea
{\rm Tr} \bSigmainv  \lefteqn{\sum_i  \left( \bGi - \bGbar + \bGbar - \bGm \right)
               \left( \bGi - \bGbar + \bGbar - \bGm \right)^T}  \nonumber \\
   & = & 
      {\rm Tr} \bSigmainv \sum_i  \left( \bGi - \bGbar \right)
                                  \left( \bGi - \bGbar \right)^T \hspace{2cm} \\
&  &+ n {\rm Tr} \left( \bGbar - \bGm \right)^T  \bSigmainv
                       \left( \bGbar - \bGm \right). \nonumber
\eea
The first term is independent of the model $A$ and contributes only to the normalization, thus
\be
    P(G | A) \propto e^{-\f{1}{2} n {\rm Tr} \left( \bGbar -\bGm \right)^T \bSigmainv \left( \bGbar -\bGm \right)}.
\ee
Note the explicit factor of $n$ which makes the distribution more peaked as it increases.
Thus for more accurate QMC data (larger $n$) a spectral function that fits the data well becomes increasingly more likely than one that does not fit so well. 
This factor of $n$ reflects the well known fact that the variance of the mean value is down by a factor $1/n$. The value of $n$ is of course rather meaningless without also specifying the number of measurements $N_{\rm meas}$ in each QMC bin, which determines the magnitude of the components of $\Sigma$. However, for a fixed large enough value of $N_{\rm meas}$, $\Sigma$ is largely independent
of $n$, thus the explicit factor of $n$ reflects accurately how the likelihood function sharpens up when more measurements of QMC data is made.

\section{Monte Carlo implementation \label{Monte_Carlo_implementation}} 
The task of sampling the posterior distribution can be done efficiently using a Monte Carlo simulation that samples the distribution $P(A)e^{-\kappa E(A)}$. $P(A)$ is the prior probability, and the energy $E(A)$ comes from the likelihood function and is  
\be
E(A) = \f{1}{2} n {\rm Tr} \left( \bGbar -\bGm \right)^T \bSigmainv \left( \bGbar -\bGm \right),
\ee
and $\kappa=1$. 

In devising a Monte Carlo procedure one can choose the probability of accepting a new spectral function $A^\prime$ as
\be \label{accept_probability}
 p(A \to A^\prime) = P(A^\prime) {\rm min}(1,e^{-\kappa(E(A^\prime)-E(A))}).
\ee
To implement the prior probability $P(A)$ according to Eq.~(\ref{generalprior}) one starts with a spectral function that is positive everywhere and satisfies the sum rule. In subsequent Monte Carlo moves one simply does not accept spectral functions which violate the positivity and the sum rule. Thus $P(A)$ is unity for allowed spectral functions and zero otherwise. Typically a simulation is started with all spectral weight concentrated at one frequency. In a Monte Carlo move spectral weight is shared between neighboring frequencies in the following manner. First a pair of neighboring frequencies $\omega_i$ and $\omega_{i+1}$ are chosen at random, and the contribution to the zero-moment sum rule from the spectral weights at these frequencies are computed: $c_0=A_{\omega_i} K_{0 \omega_i} + A_{\omega_{i+1}} K_{0 \omega_{i+1}}$. Then a random number $r$ is selected in the interval $[-c_0,c_0]$, and new spectral weights 
\bea
  A^\prime_{\omega_i} & = & A_{\omega_i} + r K_{0\omega_{i+1}}/(K_{0\omega_i}+K_{0\omega_{i+1}}) \nonumber \\
  A^\prime_{\omega_{i+1}} & = & A_{\omega_{i+1}} - r K_{0\omega_{i}}/(K_{0\omega_i}+K_{0\omega_{i+1}}) 
\eea
are proposed. Note that the zero moment sum rule is unchanged as $A_{\omega_i} K_{0 \omega_i} +A_{\omega_{i+1}} K_{0 \omega_{i+1}} = A^\prime_{\omega_i} K_{0 \omega_i} +A^\prime_{\omega_{i+1}} K_{0 \omega_{i+1}}$.
This proposed move is accepted with the probability specified in Eq.~(\ref{accept_probability}). In particular, if either of the $A^\prime$s are negative the proposed move is rejected. Note that for detailed balance to hold in this scheme $c_0$ must not change in a Monte Carlo move. For closely spaced frequencies this Monte Carlo move has a good acceptance rate. To further ensure that the simulation does not get stuck in a local energy minimum we combine this move with a parallel tempering scheme in which several simulations of the system is simultaneously carried out at different temperatures $1/\kappa$ and a swapping move between different temperature configurations is included. In order to optimize the list of temperatures we have used the scheme in Ref.~\onlinecite{Katzgraber} where the maximum movement of configurations from the highest to the lowest temperatures is achieved.

In Ref.~\onlinecite{Stocon} it was suggested that the entropy of the averaged spectrum be plotted vs. $\kappa$ and the final spectrum would be selected as the average at a value of $\kappa$ just before the entropy makes a final drop at high values of $\kappa$. We do not adopt such a procedure here as we find it undesirable to have a procedure for selecting the spectral function that depends on properties of the spectral function itself. Even though a high value of $\kappa$ gives solutions close to the most probable one there are no guarantees that the correct spectrum will not have a low entropy as is the case if the spectrum is well approximated by a single or a few narrow peaks. A similar criterion was proposed in Ref.~\onlinecite{Beach} where the value of $\kappa$ corresponding to a jump in the specific heat was chosen.

Instead we take the point of view that the final answer is the average spectrum at $\kappa=1$, which corresponds to the posterior distribution\cite{White}. This means that the resulting spectrum will depend on the accuracy of the input data, $n$. This is advantageous as it provides a mechanism against over-interpreting low quality data. However, it also means that one needs to monitor how larger values of $n$ will influence the final result. Thus a convergence analysis with $n$ is required. This makes the method rather dependent on efficient QMC algorithms as generally large values of $n$ are needed.

\section{Applications \label{Applications}}
For neutron scattering the experimentally relevant measured quantity is the dynamic structure factor
\be
   S^{ij}_q(\omega) = \int_{-\infty}^{\infty} dt e^{i \omega t} \langle S^i_q(t) S^j_{-q}(0) \rangle
\ee
where the superscripts $i,j$ indicate spin polarization directions being either $x$,$y$ or $z$, and $S^i_q(t)$ is the $i$'th polarization component of the spin operator in the Heisenberg representation at momentum $q$. For convenience we will choose units such that the lattice spacing is one.
In QMC the accessible counterpart to the dynamic structure factor is the imaginary time correlation function
\be
   {\tilde S}^{ij}_q(\tau) = \langle S^i_q(\tau) S^j_{-q}(0) \rangle.
\ee
Using the Lehmann representation one finds that $S^{ij}$ and ${\tilde S}^{ij}$ are related by
\be
   {\tilde S}^{ij}_q(\tau)= \int_0^{\infty} \f{d\omega}{2\pi} \left( e^{-\omega \tau} + e^{-(\beta-\tau)\omega} \right) S^{ij}_q(\omega),
\ee
where $\beta$ is the inverse temperature. 
Thus the kernel $K_{\tau \omega}$ in Eq.~(\ref{discreteeq}) is
\be
    K_{\tau \omega} = 
    \left\{ 
    \begin{array}{ll}
      \f{1}{2\pi}  & ,\omega = 0 \\
      \f{1}{2\pi}  \left( e^{-\omega \tau} + e^{-(\beta-\tau)\omega} \right)
      & ,\omega \neq 0.
    \end{array}
    \right.
\ee

\subsection{Antiferromagnetic dimer in a magnetic field}
In order to check the suitability of the ASM for finding the spectral function we do a test on a simple system with a non-trivial spectrum having two peaks. We choose the trivial Hamiltonian of two spins in a magnetic field $B$
\be
   H = J \vec{S}_1 \cdot \vec{S}_2 - B (S^z_1+S^z_2).
\ee
The dynamic structure factor of the transverse field components $S^{xx}_\pi(\omega)$ displays delta-function peaks at $\omega = J \pm B$ each of weight $\pi/[4(1+e^{-\beta J}(1+2\cosh \beta B))]$ which becomes $\pi/4$ at low temperatures.

We simulated this two-spin Hamiltonian at an inverse temperature $\beta J=10$ using the stochastic series expansion QMC method\cite{SSE} with directed loop updates\cite{SS}. In the simulations we extracted the imaginary time correlation function in the $x$-direction at momentum vector $\pi$. 
The imaginary time data were obtained on an equally spaced grid with 101 points from 0 to $\beta/2$, and the relative error of the imaginary time data ranged from $\sim 10^{-5}$ at small $\tau$ to $\sim 10^{-2}$ at $\tau=\beta/2$.    
The imaginary time data was then used as input to the ASM program where we used a regular grid with 200 frequencies having spacing $\Delta \omega=0.01J$.
\begin{figure}
\includegraphics[clip,width=8cm]{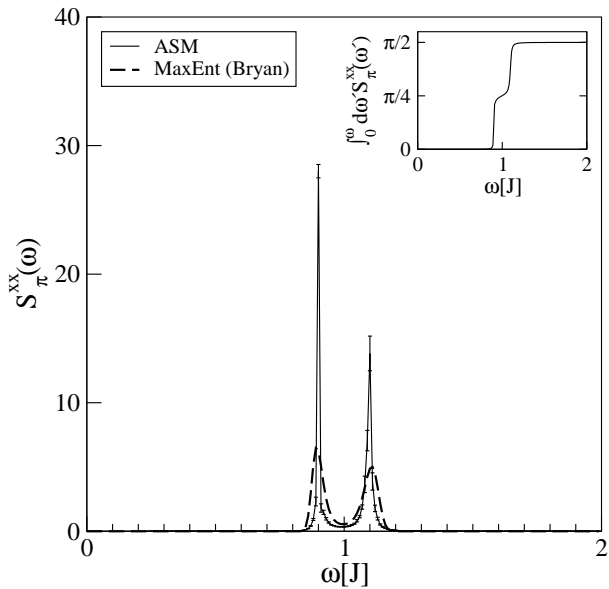}
\caption{ 
Real-frequency dynamic structure factor $S^{xx}_\pi(\omega)$ obtained from ASM (solid line) and MaxEnt (dashed line) for the two-spin Hamiltonian. The magnetic field value $B/J=0.1$. The inset shows the integrated spectrum for the ASM curve. 
\label{dimer}} 
\end{figure}

\begin{figure}
\includegraphics[clip,width=8cm]{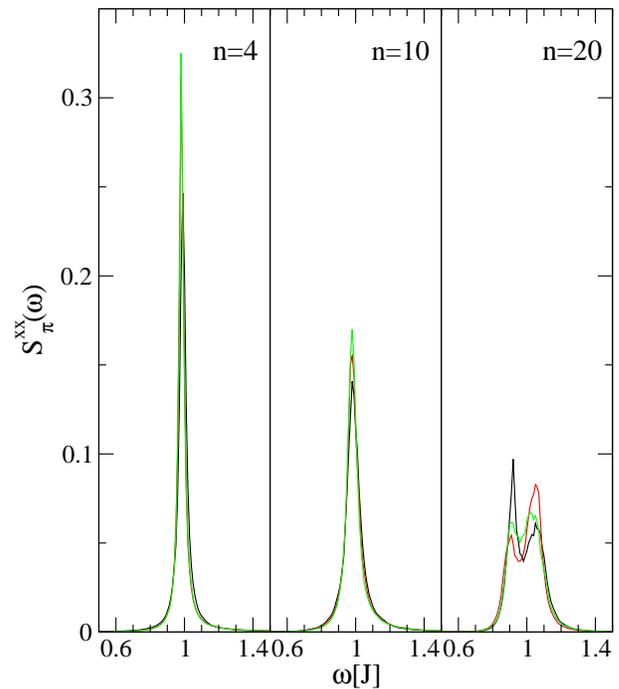}
\caption{(Color online) 
The effect of improving the data quality by increasing the number of Monte Carlo bins $n$. Each panel shows the dynamic structure factor for $B=0.1J$ for three independent data set (different line styles). The number of data bins were $n=4$ (left), $n=10$ (middle), and $n=20$ (right). For comparison the results shown in Fig.~\ref{dimer} was carried out using $n=200$.  
\label{MAGNDIMER1}} 
\end{figure}

The results for the magnetic field value $B/J=0.1$ is shown in Fig.~\ref{dimer}. This result is compared to the spectrum obtained from the same QMC data using Bryan's MaxEnt method. All methods using the entropic prior gives a possibility of including a default model so that the entropy is maximized when the spectral function matches the default model. We have used a flat model here as that corresponds most closely to our ASM choice of putting in minimal prior information. The curves in Fig.~\ref{MAGNDIMER1} were obtained using codes based on Ref.~\onlinecite{MaxEntReview}.  

From Fig.~\ref{dimer} we see that both methods are able to resolve the peaks even though the separation $2B/J = 0.2$. The peak locations corresponds well to the true value for both methods, but the ASM peaks are a bit narrower than the MaxEnt peaks. 

While the ASM gives rather sharp peaks, the two peaks are not equal as dictated by the exact solution. There is a tendency that the high energy peak is lower and broader than the low energy peak. This is also seen for the MaxEnt peaks. The spectral weight is however equally distributed on the two peaks in both the low and the high field cases, see inset of Fig.~\ref{dimer}. We expect that the peaks become more and more equal as the quality of the QMC data is increased (larger $n$).  This has the effect that the likelihood function becomes more peaked and more details of the spectrum will be better resolved. An example of this is shown in Fig.~\ref{MAGNDIMER1} where it is clear the the double peak structure is only revealed for data of sufficient quality.  

We have also simulated the dimer system with a bigger value of the magnetic field, B=0.4J. For this value of $B$ the peaks at $\omega=0.6J$ and $\omega=1.4J$ are very narrow in both the ASM and the Bryan MaxEnt method. 

\subsection{Spin-1 chain}
\begin{figure}
\includegraphics[clip,width=8cm]{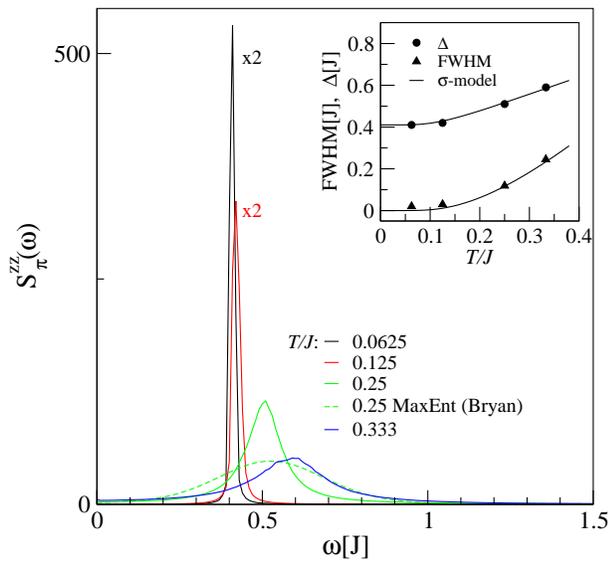}
\caption{(Color online) 
Dynamic structure factor $S^{zz}_\pi(\omega)$ for a 1D spin chain with $64$ sites obtained from the ASM (solid lines) at different temperatures indicated by the legends. The dashed curve is the MaxEnt result for $T/J=0.25$. The curves for $T/J=0.0625$ and $T/J=0.125$ have been scaled down by a factor $1/2$ to fit inside the figure boundaries. The inset shows the peak positions $\Delta$ (circles) and peak widths FWHM (triangles) as functions of temperature. The solid lines are the $\sigma$-model predictions for these quantities. 
\label{pipeak}} 
\end{figure}

We now move on to a nontrivial example, the spin-1 antiferromagnetic chain, the so called Haldane chain. The Haldane chain is famous for being gapped in contrast to the half-integer spin chains\cite{Haldane}. The minimum gap is at $Q=\pi$ in units of the inverse lattice spacing. Fig.~\ref{pipeak} shows plots of $S^{zz}_{Q=\pi}(\omega)$ for different temperatures obtained using the ASM. Note how the peak position and width increase with temperature. To compare with MaxEnt we have shown the MaxEnt result using Bryan's method for a single temperature $T/J=0.25$ as a dashed curve. Note that the MaxEnt curve captures the peak position well, but gives a very broad peak.  The inset shows a comparison of the temperature dependence of the gap vs. a non-linear sigma model prediction which was obtained by solving the finite temperature gap equation in Ref.~\onlinecite{Jolicour} numerically.
In the inset we also show a comparison of the width of the peaks, quantified by their full width at half maximum (FWHM), with predicted values from a combined nonlinear $\sigma$-model and scattering matrix calculation\cite{Damle}. The agreement is quite remarkable and involves no adjustable parameters. 

\begin{figure}
\includegraphics[clip,width=8cm]{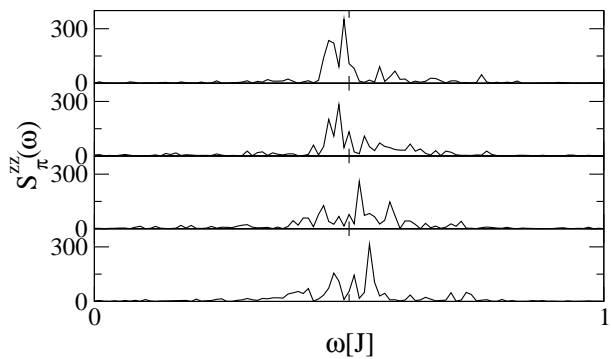}
\caption{ 
Snapshots of spectra. These spectra (and others) are averaged over in order to yield the result shown in Fig.~\ref{pipeak}. The spectra here are all for $T/J=0.25$. 
\label{snapshots}} 
\end{figure}

One can ask whether the temperature broadening of the peak seen in Fig.~\ref{pipeak} obtained using the ASM is just due to the ``motion'' of a single sharp peak. Fig.~\ref{snapshots} shows that this is not the case.

\subsection{Bond alternating antiferromagnetic chain}
Another nontrivial spin model is the bond alternating spin-1/2 Heisenberg chain (BAHC) which has been studied extensively, and is relevant for materials such as ${\rm Cu(NO}_3{\rm)}_2\cdot{\rm 2.5D}_2{\rm O}$\cite{CUNODO0,CUNODO1,CUNODO2}, the spin-Peirels material ${\rm CuGeO_3}$\cite{CuGeO3} and others (See Ref.~\cite{Barnes}). 
The Hamiltonian for the BAHC is
\be \label{BAHChamiltonian}
   H = J \sum_i \left( \vec{S}_{2i-1} \cdot \vec{S}_{2i} + 
                       \lambda \vec{S}_{2i} \cdot \vec{S}_{2i+1} \right) 
\ee
where $\lambda \geq 0$.      
Although the BAHC is a one-dimensional model, it is not solvable by the Bethe Ansatz. Thus other techniques are needed to obtain the dynamics. In this regard investigations using bosonization\cite{Tsvelik} the RPA approximation\cite{UhrigSchulz}, series expansions\cite{GelfandSingh,Barnes,Trebst,Weihong,Hamer,Zheng,Singh} and exact diagonalization studies\cite{Mikeska} have produced very impressive results for the dynamics of the BAHC containing predictions of the dispersion of one magnon excitations as well as bound states and details about multi particle excitations.

\begin{figure}
\includegraphics[clip,width=8cm]{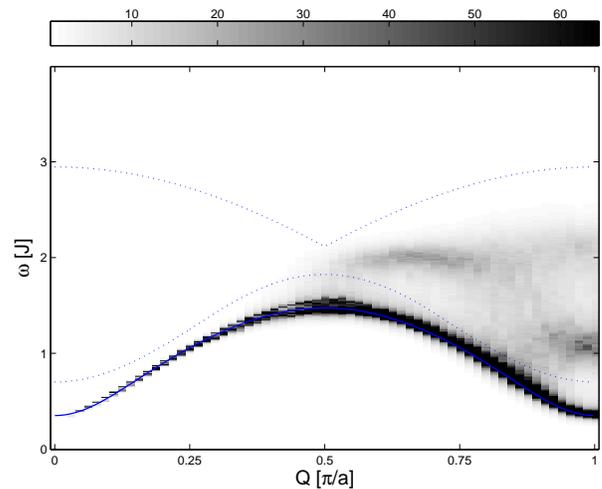}
\caption{ 
Gray scale plot of $S^{zz}_Q(\omega)$ for the BAHC with $\lambda=0.8$. The simulations were carried out at $\beta J=16$ on a lattice with 128 sites and periodic boundary conditions. The solid blue curve indicates the one-magnon excitations as calculated using a series expansion about the dimer limit\cite{Barnes}, and the dotted lines show the kinematic boundaries of two-particle excitations.
\label{bahc}} 
\end{figure}

We carried out QMC simulations of the BAHC for a chain with 128 sites and periodic boundary conditions at inverse temperature $\beta J=16$ and $\lambda=0.8$. The ASM was used to obtain the spectra at all momentum points.
Figure \ref{bahc} shows a gray scale plot of $S^{zz}_Q(\omega)$ for different values of  $Q$ and $\omega$ . The one magnon excitations are easily identified as the sharp dark feature and agrees very well with that obtained from series expansion to order $\lambda^5$,\cite{Barnes} shown as the blue solid curve. For $Q \gtrsim 0.5\pi$ many-particle excitations are visible. This agrees qualitatively with the results in Ref.~\onlinecite{Singh} which shows that the many-particle continuum has appreciably more spectral weight for $Q \gtrsim 0.5\pi$ than for smaller $Q$. 
For $0.5\pi \lesssim Q \lesssim 0.75\pi$ there is an almost flat feature in the continuum at $\omega \sim 1.9J$ which is well separated from the band of one magnon excitations and also from the kinematic boundaries of two magnon excitations shown as blue dotted lines. This is not seen from the series expansion\cite{Singh} and RPA results\cite{UhrigSchulz} which predict that the continuum should have biggest spectral weight at its lower boundary. However, this feature is reminiscent of that seen in experiments on ${\rm Cu(NO}_3{\rm)}_2\cdot{\rm 2.5D}_2{\rm O}$\cite{CUNODO2} where a dispersion-less feature in the continuum was reported. As $Q$ is increased towards $\pi$ this feature broadens and vanishes. Some structure reappears in the continuum close to $Q=\pi$ where a peak at $\omega \sim J$ and a very weak feature at $\omega \sim 2J$ is seen. 
A word of caution is needed in interpreting weak features of Fig.~\ref{bahc}. This is because Fig.~\ref{bahc} also shows occurrence of spectral weight in between the one magnon peak and the lower kinematic boundary of the two magnon excitations, where one expects a gap. This is probably caused by insufficient quality of the QMC data which gives spectral weight in unwanted places in a similar fashion to what is seen in Fig.~\ref{dimer} at $\omega \sim J$ for $B=0.1J$.

\begin{figure}
\includegraphics[clip,width=8cm]{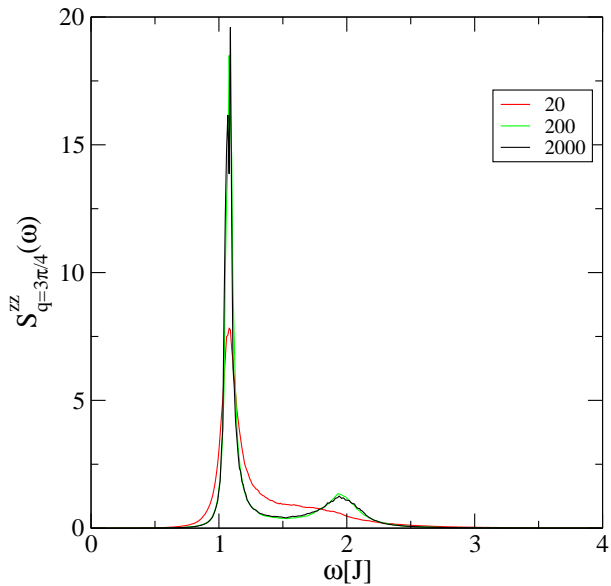}
\caption{ 
Line shapes at a fixed momentum $Q=3\pi/4$ for QMC data sets of different lengths $n$ indicated by the legends.
\label{altchains075}} 
\end{figure}

The QMC data plotted in Fig.~\ref{bahc} were taken from a run with in all $n=2000$ data bins. In order to see how the number of QMC bins affect the line shapes we show in Fig.~\ref{altchains075} the dynamic structure factor at $Q=3\pi/4$ for three different values of $n$. While there is some significant change in the line shape from $n=20$ to $n=200$, increasing $n$ to 2000 has only minor effects.

\begin{figure}
\includegraphics[clip,width=8cm]{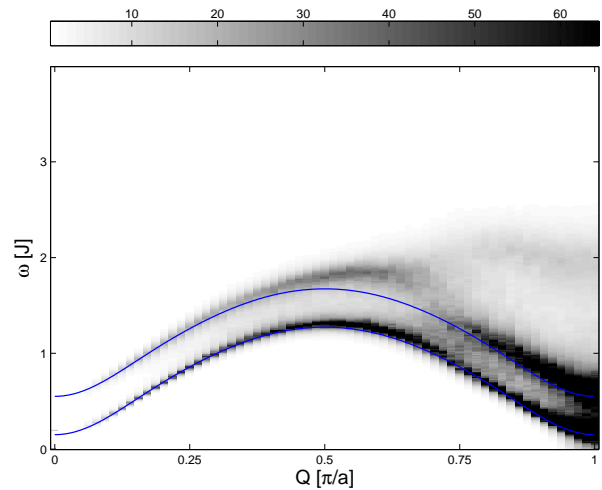}
\caption{ 
Gray scale plot of $S^{xx}_Q(\omega)$ for the BAHC with $\lambda=0.8$ in a magnetic field $B=0.2J$. The inverse temperature is $\beta J=16$ and $L=128$. The solid lines are the spin-split one magnon result. 
\label{altfield02}} 
\end{figure}

We will now add a magnetic field term $-B\sum_i S^Z_i$ to Eq.~(\ref{BAHChamiltonian}). For $\lambda=0$ the BAHC is just a collection of independent antiferromagnetic dimers. When subjecting a dimer to a magnetic field in the spin z direction the degeneracy of the spin triplet excitations is lifted, and one expects a double-peak structure, as seen in Fig.~\ref{dimer}, in the transverse dynamic structure factor $S^{xx}$. For finite $\lambda$ the dimers become coupled, however one still expects the splitting to occur, at least for small values of the magnetic field. Fig.~\ref{altfield02} shows a gray scale plot of $S^{xx}_Q(\omega)$ for $\lambda=0.8$ and a small value of the magnetic field $B=0.2J$. The splitting of the one magnon peak is clearly seen and agrees, for small values of $Q$, well with the expectation that the effect of the magnetic field is simply to displace the one magnon dispersion by $\pm B$. The solid lines indicate this. We have taken the one magnon dispersion from the series expansion\cite{Barnes} and added(subtracted) an energy $B=0.2J$. For $0.5\pi \lesssim  Q \lesssim 0.75\pi$ there are deviations from this simple picture, as the upper branch is higher in energy and broadens considerably. For even higher momentum values there is significant broadening of the peaks and at $Q=\pi$ they are hardly distinguishable. For $Q \gtrsim 0.75\pi$ one can also see the appearance of many-particle excitations above the one magnon peaks.     

\begin{figure}
\includegraphics[clip,width=8cm]{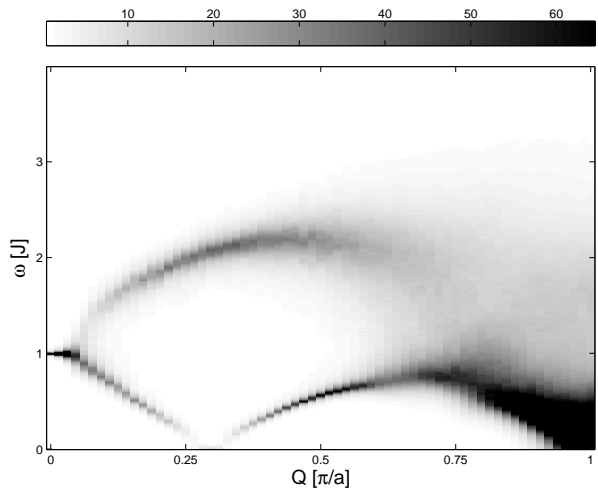}
\caption{ 
Gray scale plot of $S^{xx}(Q,\omega)$ for the BAHC in a magnetic field $B=J$. $\lambda=0.8$, $\beta J=16$ and $L=128$.  
\label{altfield10}} 
\end{figure}

For a large value of the magnetic field the lower branch goes to zero energy at a certain characteristic value of the momentum. Figure \ref{altfield10} shows a gray scale plot of the transverse structure factor $S^{xx}_Q(\omega)$ for $\lambda=0.8$ and $B=J$. One can clearly see that there is a branch of excitations that approaches zero at $Q \approx 0.3\pi$ and at $Q=\pi$. This is consistent with the results reported in Ref.~\onlinecite{Chitra}. It is also apparent that the intensity at $Q \approx 0.3\pi$ vanishes as the energy approaches zero, while the intensity at $Q=\pi$ is high. The high energy magnon branch is clearly seen for $Q \lesssim 0.6\pi$ and gets broadened considerably and disappears for larger $Q$. There is also a sharp finite energy peak seen at small $Q$ resulting from the merger of the two magnon branches. 

\subsection{Heisenberg antiferromagnetic chain}
The spin-1/2 Heisenberg chain was the first nontrivial quantum many-body problem to be solved exactly\cite{Bethe}. Yet it is still only recently that exact results for the dynamical correlation functions have appeared\cite{Caux}. We compare here the ASM with the exact numerical result for the dynamic structure factor for the Heisenberg antiferromagnetic chain.

\begin{figure}
\includegraphics[clip,width=8cm]{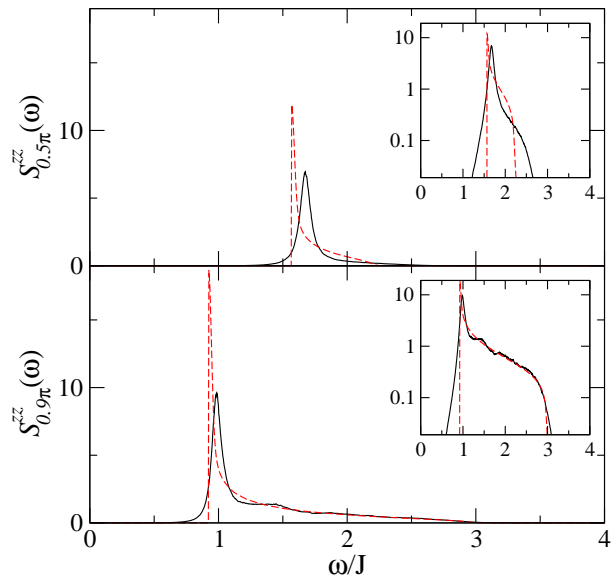}
\caption{(Color online) Line shapes of $S^{zz}(Q,\omega)$ for the 1D Heisenberg antiferromagnet at $Q=0.5\pi$ (upper panel) and $Q=0.9\pi$ (lower panel) solid curves. The red dashed lines are exact results obtained from the Bethe Ansatz. The chain has periodic boundary conditions and has $L=500$ sites. The QMC simulations are carried out at $\beta J=40$ while the Bethe Ansatz result is obtained at $T=0$. The insets shows the same results but on a semi-log scale.
\label{heisenbergchain}} 
\end{figure}

In Fig.~\ref{heisenbergchain} we show the lineshape of $S^{zz}(Q,\omega)$ at $Q=0.5\pi$, where the gap is the largest, as well as at $Q=0.9\pi$ where the exact result has a very long high-energy tail. We see that the exact results (red dashed curves) is zero up to a certain energy where a vertical leading edge marks the onset of a continuum of excitations. The ASM results have no true vertical leading edge, but rather a power-law increase. This smooth increase is inevitable in the ASM method as even a prior that incorporates a strict requirement of having a vertical leading edge will give a smooth leading edge if there is uncertainity about the position of the edge. There is also a slight difference in the location of the maximum intensity. While the exact results peak right at the leading edge, the ASM results peak slightly above the exact results. This is most prominent in the $Q=0.5\pi$ case and is probably because the true lineshapes are very asymmetric and tend to push up the peak in energy. This asymmetry can also be seen in both ASM curves. The extent of how high up in energy the continuum reaches can be seen from the insets. The high energy tail is very well reproduced by the ASM for $Q=0.9\pi$ while it is overestimated for the $Q=0.5\pi$ case.  

\subsection{Square lattice Heisenberg antiferromagnet}
The spin-1/2 square lattice Heisenberg antiferromagnet (2DAF) has been studied intensively because of its relevance to the cuprate materials that are superconducting at high temperatures when doped. The dynamics of the 2DAF is rather well described by linear spin-wave theory\cite{LSW}. However, linear spin wave theory does not account for a magnon dispersion along the zone boundary. Such a dispersion was predicted using an expansion around the Ising limit\cite{SeriesIsing,Zheng} and indicates a difference in energy between the magnon peaks at $(\pi,0)$ and $(\pi/2,\pi/2)$ of about 7-9\%, the energy at $(\pi/2,\pi/2)$ being the highest. Similar result was obtained using QMC: In Ref.~\onlinecite{SandvikSingh} the QMC data were fitted to a functional form consisting of a delta-function and a broad continuum, while in Ref.~\onlinecite{Ronnow} the MaxEnt method was used. Higher order Holstein-Primakoff spin wave calculations gives a smaller value, 2\% \cite{Igarashi2}, as does an expansion based on the Dyson-Maleev transformation\cite{Canali,Canali2}.

\begin{figure}
\includegraphics[clip,width=8cm]{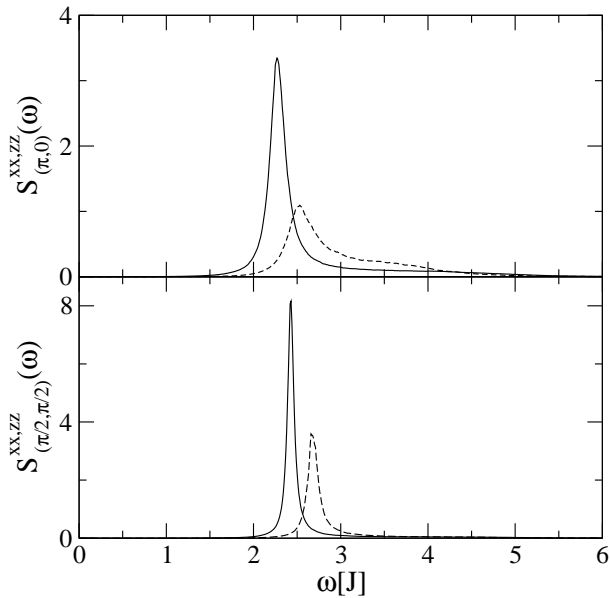}
\caption{ 
Transverse (solid curves) and longitudinal (dashed curves) dynamic structure factor for the 2DAF at $Q=(\pi,0)$ (upper panel) and $Q=(\pi/2,\pi/2)$ (lower panel).
\label{S2DAF}} 
\end{figure}

 Experimental measurements of the material copper formate tetradeuterate(CFTD)\cite{CFTD,CFTD2} indicated a difference of 7\% in agreement with the series expansion results and the QMC, however ${\rm La}_2{\rm CuO}_4$ shows\cite{Coldea} an entirely different dispersion with the peak at $(\pi,0)$ being higher in energy than at $(\pi/2,\pi/2)$. This dispersion has been explained as special features of the Hubbard model\cite{Peres}. Recently experiments on ${\rm K}_2{\rm V}_3{\rm O}_8$,  also supposedly a realization of the Heisenberg antiferromagnet on the square lattice, showed a double peak structure of unknown origin at $(\pi/2,\pi/2)$\cite{Lumsden}. In order to investigate this possible double peak structure we repeated the simulations of Ref.~\onlinecite{SandvikSingh} and analyzed the imaginary time data using the ASM which gives unbiased information about the line shapes. In order to distinguish transversal and longitudinal excitations the simulations were carried out as in Ref.~\onlinecite{SandvikSingh} by imposing a staggered magnetic field $H_{\rm stag}=0.001615$ that yields a staggered magnetization consistent with the experimental value $m_s=0.307$ on a $32 \times 32$ lattice at an inverse temperature $\beta J=32$. We measured both the transverse dynamic structure factor  $S^{xx}$ and the longitudinal one $S^{zz}$. The results for the two momentum points $Q=(\pi,0)$ and $Q=(\pi/2,\pi/2)$ are shown in Fig.~\ref{S2DAF}. We observe a difference in magnon energies in the transverse channel corresponding to  $(E_{(\pi/2,\pi/2)}-E_{(\pi,0)})/E_{(\pi/2,\pi/2)} \approx 6\%$, determined from the location of the maximum. However, the peak locations are at slightly higher energies than the corresponding delta-function locations found in Ref.~\onlinecite{SandvikSingh}.  As we expect a priori that the dynamic structure factor in the transverse channel contains a delta-function like one magnon peak and a continuum we believe that the result in Ref.~\onlinecite{SandvikSingh} is the most accurate as it accounts for more prior information. However, for the longitudinal channel the expected functional form of the spectral function is not so clear. In particular it is not obvious that the particular functional form chosen in Ref.~\onlinecite{SandvikSingh} in the longitudinal channel is flexible enough to track the real line shape. In fact, in contrast to the result reported there, at $Q=(\pi/2,\pi/2)$, the lower panel of Fig.~\ref{S2DAF} shows that the peak location in the transverse channel is at a substantial lower energy $(\sim 10\%)$ than the peak in the longitudinal channel.  For an experiment that measures both the longitudinal and transverse structure factors simultaneously this could give rise to a double peak structure at $(\pi/2,\pi/2)$.  Such a double peak should also be apparent at $(\pi,0)$, although more weakly, because the longitudinal structure factor is more strongly peaked at $(\pi/2,\pi/2)$ than at $(\pi,0)$. In fact, as can be seen from Fig.~\ref{S2DAF} the longitudinal dynamic structure factor at $(\pi,0)$ has a very long high-energy tail.   

\section{Summary \label{Summary}}
Obtaining equilibrium dynamics from numerical imaginary time correlation functions is an important task. We have in this article investigated the suitability of a specific Bayesian method for doing this.
This method, known as the ASM\cite{White}, proposed already in 1991 has not been widely used. We suspect that this is because its nature is such that for it to give good results one needs rather accurate QMC data. However, QMC simulations have improved considerably the last years, thus it is timely to reconsider its usefulness. The ASM is a Bayesian data analysis method, where instead of picking the final result as the spectrum that maximizes the posterior probability distribution, the final answer is picked as the averaged spectrum over the posterior probability distribution. The reason for selecting to take the average is the rather unselective nature of the specific prior probability distribution used. We argue for the use of a prior probability distribution that encodes just hard knowledge; spectral positivity and sum rules, and the specific form of the prior is then the one maximizing the information theory entropy under these constraints. One should note that this prior is not the entropic prior used in various MaxEnt methods. The entropic prior gives high probabilities to spectral functions that itself has high entropy, thus favoring smooth spectral functions. 

There are other methods that resembles the ASM. The Stochastic continuation method\cite{Stocon} is essentially the same method, except for the use of a drop in entropy as the criterion for determining the temperature at which the sampling is carried out. In the ASM the posterior probability distribution is sampled directly. Thus in essence the quality of the input data determines the effective sampling temperature which is implicit in the approach. We find this desirable as it protects from over interpreting bad data and makes the procedure independent of the particular form of the spectral function itself. However, this also implies the need of a convergence analysis of the obtained spectral function with increasingly better QMC data. Some MaxEnt methods, such as the Bryan MaxEnt method, also outputs as the final answer an averaged spectrum. In the case of Bryan's method\cite{Bryan} the average is taken over the probability distribution of the coefficient determining the relative importance of the entropic prior.

The ASM is on at least as firm statistical footings as other Bayesian methods\cite{MaxEntReview}. It has the disadvantage of being computationally demanding, however it is not as computer-intensive as the QMC simulations themselves. A typical run of the ASM, for one momentum space point, takes about 4 hours on an Intel Pentium IV, 2.4 GHz processor. In comparison running MaxEnt methods takes typically of the order of tens of seconds. 

In showing examples of the ASM we have sampled the posterior probability distribution and obtained spectral functions for several model systems. Of new results we have shown that using this method we can obtain the finite temperature position and broadening of the Haldane gap in spin-1 antiferromagnets, and that the results agree very well with nonlinear $\sigma$-model predictions without any adjustable parameters. We have also applied the method to the spin-1/2 Heisenberg chain with alternating bond strengths where we found a quantitative very good agreement with other methods for the dispersion of one magnon excitations. We also observed some structure in the continuum of many-particle excitations which have not been seen using other methods. At present it is unclear whether these many-particle features are real or whether they are artifacts of insufficient QMC data.
We have also added a magnetic field to the bond alternating chain, and observed the expected spin-split spectrum in the transverse dynamic structure factor. For a bigger value of the magnetic field we also see the weak incommensurate low-energy mode and the much stronger low-energy mode at $Q=\pi$. 
We have compared the ASM for the 1D Heisenberg antiferromagnet with exact Bethe Ansatz results. The comparison reveals a good similarity, although certain sharp features of the exact result, such as the lineshape's vertical leading edge, is not accurately reproduced by the ASM.  
Finally we studied the dynamic structure factor at the zone boundary for the two dimensional square lattice spin-1/2 Heisenberg antiferromagnet, and found results consistent with existing results on that system except for a difference in peak locations of the transverse and longitudinal dynamic structure factor at the same momentum value that can possibly give rise to a double peak structure in measurements using unpolarized neutrons.

\begin{acknowledgments}
The author thanks Anders Sandvik for introducing him to the Stochastic continuation method and for making him aware of Ref.~\onlinecite{White}, and Jean-S\'{e}bastien Caux for providing the Bethe ansatz results in Fig.~\ref{heisenbergchain}. The numerical simulations were in part carried out using the Nordugrid ARC middleware on SWEGRID computers provided by the Swedish National Infrastructure for Computing under the contract SNIC 021/06-64.  
\end{acknowledgments}

\end{document}